\documentclass[doublecol]{epl2} 
\usepackage{amsmath}
\usepackage{amsfonts}
\usepackage{graphicx}
\usepackage{hyperref}
\usepackage{mathtools}
\usepackage{verbatim}
\usepackage{epstopdf}
\usepackage{subfigure}
\usepackage{latexsym}
\usepackage{dcolumn}
\usepackage{epsf}
\usepackage{float}
\DeclarePairedDelimiter\abs{\lvert}{\rvert}
\DeclarePairedDelimiter\norm{\lVert}{\rVert}

\title{Optimizing synchronization in multiplex networks of phase oscillators}

\author{Prosenjit Kundu,\inst{1} Pitambar Khanra, \inst{1}  Chittaranjan Hens,\inst{2}   \and Pinaki Pal\inst{1}}

\shortauthor{Prosenjit Kundu \etal}

\institute{                    
  \inst{1} Department of Mathematics, National Institute of Technology, Durgapur 713209, India\\
  \inst{2} Physics and Applied Mathematics Unit, Indian Statistical Institute, Kolkata 700108, India
}
\pacs{05.45.Xt}{Synchronization; coupled oscillators}
\pacs{89.75.-k}{Complex systems}
 	 
\abstract{
We present an analytical scheme to achieve optimal synchronization in multiplex networks of  frustrated and non-frustrated phase oscillators.  We derive a multiplex synchrony alignment function (MSAF) for that purpose, the expression of which consists of structural as well as dynamical information of the layers of the multiplex network. Analyzing the MSAF, a set of frequencies (optimal frequencies) is determined to achieve optimal synchronization in the network. Further, using the scheme, we show that perfect synchronization can be achieved in a layer of the multiplex network for given coupling strength and phase frustration parameters. The analytical scheme presented here has been tested for heterogeneous multiplex networks of frustrated and non-frustrated Kuramoto dynamics. 
}

\begin{document}

\maketitle
Diverse collective phenomena can emerge in complex systems consisting of interacting dynamical units on complex network topology. One such emergent collective phenomenon is synchronization \cite{Pikovsky_synchronization_book,Arenas_PhysReport2008}, observed and tested in different real world systems including group of fireflies, power grid networks, brain, cellular and chemical oscillators  \cite{Strogatz_synchronization_book,Pikovsky_synchronization_book,Arenas_PhysReport2008, Motter_NatPhys2013, Belykh_PRL2005, Dorogovtsev_RMP2008}. On the other hand, significant advancement has been made in characterizing the statistical scaling of diverse complex  network topologies and its profound applications to real situations \cite{Dorogovtsev_RMP2008, Albert_RMP2002, Cohen_complex_book}. Therefore, it has become imparative to understand how the interplay between network topology and nodal states influence the emergent dynamics in complex networks\cite{Barzel_NatPhys2013,Hens_NatPhysics2019}. Researchers have been trying to interlink the collective macroscopic property such as synchronization with  network structure \cite{Arenas_PhysReport2008, Ichinomiya_PRE2004, Restrepo_PRE2005, Arenas_PRL2006, Jesus_PRL2007,Skardal_PRL2014} for long, yet it is not fully understood how structural or degree heterogeneity affects the collective emergent behavior (say synchronization) of coupled oscillators or vice versa.

Currently, multiplex network \cite{Boccaletti_PhysReport2016, Domenico_NatPhys2016, Danziger_NatPhys2019, Jalan_PRE2019} has became an interesting topic to the researchers for its diverse application in real world ranging from transportation to ecology.  For instance, the multipexity can create faster diffusion process \cite{Domenico_NatPhys2016}, promotes  synchronization in phase frustrated dynamics \cite{Kachhvah_EPL2017, Khanra_PRE2018, Nicosia_PRL2017} and leads to abrupt transition in consensus dynamics\cite{Panos_PRE2019}. However the literature lacks a detailed investigation regarding proper frequency selection in multiplex network to generate a favourable synchronization dynamics. This type of study was motivated form the emergence of explosive and perfect synchronization for degree-frequency correlated network of phase oscillators\cite{Jesus_PRL2011, Kundu_EPL2018, Brede_PRE2016}. Skardal et al.~\cite{Skardal_PRL2014} showed that  the synchronization process is easily achievable (or faster) in network of phase oscillators if the frequencies are drawn from the leading eigen vector of the Laplacian matrix. We aim here to extract a global frequency set in multiplex  network of frustrated and non-frustrated phase oscillators such that entire network synchronizes  early in comparison with other frequency  distributions. 

To start with, we consider a duplex network in which interacting oscillatory units of the individual layers are modeled by the phase oscillators\cite{Kuramoto_chemical_book}
\begin{eqnarray}
\frac{d\theta_i^{(1)}}{dt} &=& \omega_i^{(1)} + \lambda\sum_{j=1}^{{N}}A_{ij}^{(1)}\sin(\theta_j^{(1)} - \theta_i^{(1)}-\alpha)\nonumber \\ &&+  \lambda\sin(\theta_i^{(2)} - \theta_i^{(1)})\label{eqn1}\\
{\mathrm{and}}\nonumber\\
\frac{d\theta_i^{(2)}}{dt} &=& \omega_i^{(2)} + \lambda\sum_{j=1}^{{N}}A_{ij}^{(2)}\sin(\theta_j^{(2)} - \theta_i^{(2)}-\beta)\nonumber \\&& + \lambda\sin(\theta_i^{(1)} - \theta_i^{(2)}),~~ ~~i = 1\dots {N}\label{eqn2}.  
\end{eqnarray}
Here, $N$ is the total number of oscillators in each layer, $\omega_{i}^{(l)}$ is  the inherent frequency of the $i^{th}$ node of $l^{th}$ layer, matrix $A^{l}$ ($N \times N$) represents the connectivity among the units in $l^{th}$ layer of the multiplex network ($l = 1,2$) and $\lambda$ is the coupling strength. 

To quantify the coherent behavior  of all phase oscillators in layer $l$, we use Kuramoto order parameter $\mathrm{R}_{l}e^{i\psi_l} = \frac{1}{\mathrm{N}}\sum_{j=1}^{{N}}e^{i\theta_j^{(l)}}$,
and the global order parameter of the entire network (including all layers) is defined as 
$\mathrm{R}e^{i\psi} = \frac{1}{{2N}}\sum_{l}\sum_{j=1}^{{N}}e^{i\theta_j^{(l)}}$. 
Normally phase-lag parameters ($\alpha$ and $\beta$) are found to inhibit the transition to synchronization in coupled systems. Physically, presence of phase lag \cite{Lohe_Automatica2015} is very important when the synchronization is investigated in many real systems systems, namely, in the array of Josephson junctions \cite{Wiesenfeld_PRE1998}, power  network \cite{Dorfler_SIAM2012} or  in mechanical rotors \cite{Mertens_PRE2011}. 

The role of phase-lag has been investigated in~\cite{Omelchenko_PRL2012} using the theoretical Sakaguchi and Kuramoto (S-K) model~\cite{Sakaguchi_PTP1986} on complete graph. They found that the system may reveal a non-universal synchrony  for general frequency distribution. On the other hand, abrupt synchronization transition (explosive synchronization) can be captured in S-K model for a fully or partially  degree-frequency correlated  network topology \cite{Kundu_PRE2017, Kundu_Chaos2019} and a perfect synchronization (where $\mathrm{R}$ is exactly $1$) in a broad range of frustration parameter was established for a predefined coupling strength \cite{Kundu_EPL2018, Brede_PRE2016}. However, enhancement of synchronization in a multiplex network of phase-frustrated dynamics has not been explored so far.

In this paper,  we present a general analytical scheme to achieve optimal or perfect synchronization in multiplex networks of phase  (with or without phase frustration) oscillators. Main objective of this paper is to derive a  set of frequencies for which (i) $\mathrm{R}\sim 1$ for  finite  coupling strength in absence or presence of phase lag or (ii) $\mathrm{R}=1$, a scenario of perfect synchronization \cite{Kundu_EPL2018} in presence of {\it phase frustration} for arbitrary coupling strength. For this purpose, we derive a multiplex synchrony alignment function (MSAF) for multiplex networks following the approach proposed by Skardal {\it et al.}  \cite{Skardal_PRL2014}. Analyzing the MSAF we derive a set of preferable frequencies and settle  the optimality of synchronization in multiplex networks. Here, we examine  two important issues: (i)  how MSAF  determines the optimality in multiplex  network of phase oscillators and (ii) how it helps to remove the erosion effect  (by generating a desired set of frequencies) for frustrated oscillators?  Our analytical scheme describes how functional heterogeneity and structural heterogeneity or degree heterogeneity influence each other  for getting the global synchronization in phase frustrated coupled Kuramoto oscillators.

\section{Derivation: MSAF and optimal frequency set} 
In this section, first we analytically derive a multiplex synchrony alignment function (MSAF) for mutiplex networks following the approach reported in~\cite{Skardal_PRL2014}.  Then use it to derive a frequency set for achieving optimal synchronization in the network. 
 
For coherent state  ($\abs{\theta_{j}^{(l)}-\theta_{i}^{(l)}} \rightarrow 0$), 
equations~(\ref{eqn1}) and ~(\ref{eqn2}) can be written as
\begin{eqnarray}
\frac{d\theta_i^{(1)}}{dt} &=& \tilde{\omega}_i^{(1)} - \lambda\cos{\alpha}\sum_{j=1}^{\mathrm{N}}L_{ij}^{(1)}\theta_j^{(1)} + \lambda(\theta_i^{(2)} - \theta_i^{(1)}),\label{eqn3}\nonumber \\
\frac{d\theta_i^{(2)}}{dt} &=& \tilde{\omega}_i^{(2)} - \lambda\cos{\beta}\sum_{j=1}^{\mathrm{N}}L_{ij}^{(2)}\theta_j^{(2)} + \lambda(\theta_i^{(1)} - \theta_i^{(2)}), \label{eqn4}\nonumber   
\end{eqnarray}
where $\tilde{\omega}_i^{(l)}=\omega_i^{(l)}+\lambda d_i^{(l)}\sin{(-\alpha)}$ and $L_{ij}^{(l)}=\delta_{ij}d_i^{(l)}-A_{ij}^{(l)} (l = 1,2)$ and $d_i^{(l)}$ is the degree of the $i-$th node of $l$-th layer.
If the oscillators of the $l^{th}$ layer follow the synchronization manifold as $\theta_1^{(l)}\sim \theta_2^{(l)} \sim... \sim\theta_N^{(l)}\sim \psi$  we expect that the order parameters will behave as $\mathrm{R}_{l}\sim 1 (l = 1, 2)$. On the other hand, $\mathrm{R}_{l}$ will be zero if the oscillators are randomly distributed to the circumference of the unit circle.
\begin{figure*}
\includegraphics[height=!,width=0.95\textwidth]{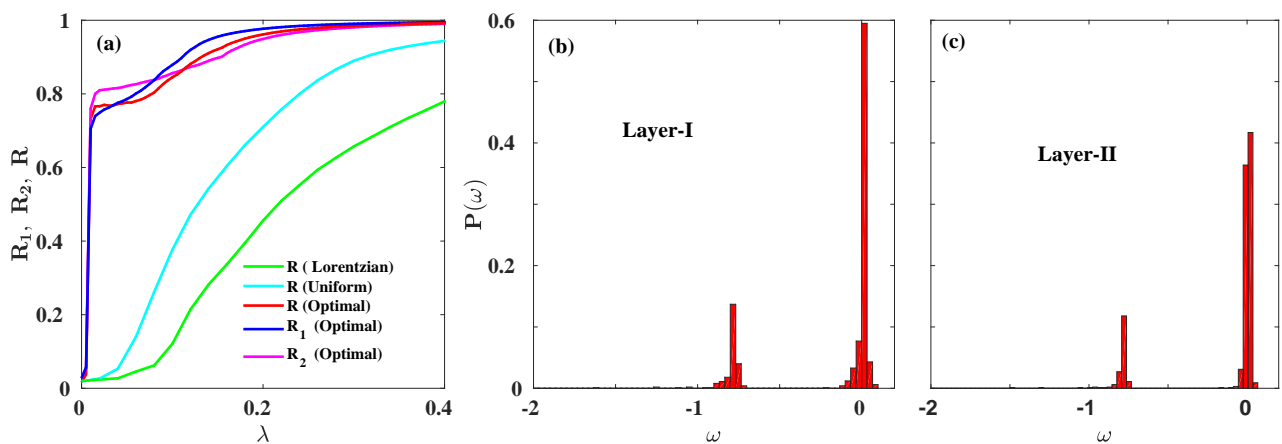}
\caption{{\bf Optimal synchronization in multiplex network of non-frustrated phase oscillators.} (a) Global order parameter ($\mathrm{R}$) as a function of coupling strength ($\lambda$) for a multiplex network with two layers in which layer-I is a scale-free network of size  $N= 1000$ with mean degree $\langle k_1 \rangle =8$  and layer-II is a scale-free network of same size and mean degree $\langle k_1 \rangle =12$. The blue, red and sky-blue line respectively represent the global order parameter computed for the network using the proposed distribution, Lorentzian and uniform distribution of $\boldsymbol{\omega}$.  Optimal frequency distributions for layer-I (b) and layer-II (c).} 
\label{fig:1}
\end{figure*}
Now in vector form the equations can be written as 
\begin{eqnarray}\label{eqn3}
\dot{\boldsymbol{\theta}}^{(1)} &=& \tilde{\boldsymbol{\omega}}^{(1)} - \lambda\cos{\alpha}L^{(1)}\boldsymbol{\theta}^{(1)} + \lambda(\boldsymbol{\theta}^{(2)} - \boldsymbol{\theta}^{(1)}),\\
\dot{\boldsymbol{\theta}}^{(2)} &=& \tilde{\boldsymbol{\omega}}^{(2)} - \lambda\cos{\beta}L^{(2)}\boldsymbol{\theta}^{(2)} + \lambda(\boldsymbol{\theta}^{(1)} - \boldsymbol{\theta}^{(2)}),  
\end{eqnarray}
where $L^{(l)} = (L_{ij}^{(l)})_{N\times N} (l = 1,2)$.

Now in steady state, with a proper choice of reference frame we have $\dot{\boldsymbol{\theta}}^{(1)}=\dot{\boldsymbol{\theta}}^{(2)}=0$.
Therefore, 
\begin{eqnarray}\label{theta_fp}
{\boldsymbol{\theta}^{(1)}}^* &=& \frac{1}{\lambda} {L_{m}^{1}}^\dagger \tilde{\boldsymbol{\omega}}_m^{(1)}, \\
{\boldsymbol{\theta}^{(2)}}^* &=& \frac{1}{\lambda} {L_{m}^{2}}^{\dagger} \tilde{\boldsymbol{\omega}}_m^{(2)}
\end{eqnarray}
where 
\begin{eqnarray}
L_{m}^{(1)} &=& \cos \alpha L^{(1)} +\cos \beta L^{(2)} +\cos\alpha \cos\beta L^{(2)}L^{(1)}, \label{lm1}\\
L_{m}^{(2)} &=& \cos \alpha L^{(1)} +\cos \beta L^{(2)} +\cos\alpha \cos\beta L^{(1)}L^{(2)},\\ \label{lm2}
\tilde{\boldsymbol{\omega}}_m^{(1)}&=&\tilde{\boldsymbol{\omega}}^{(1)}+\tilde{\boldsymbol{\omega}}^{(2)} +\cos \beta L^{(2)}\tilde{\boldsymbol{\omega}}^{(1)},\\\label{omega_modified1}
\tilde{\boldsymbol{\omega}}_m^{(2)}&=&\tilde{\boldsymbol{\omega}}^{(1)}+\tilde{\boldsymbol{\omega}}^{(2)} +\cos \alpha L^{(1)}\tilde{\boldsymbol{\omega}}^{(2)}.\label{omega_modified2}
\end{eqnarray}
The order parameter of the $l^{th}$ layer then can be rewritten as 
\begin{eqnarray}\label{op_SAF}
R_{l} &=& 1-\frac{1}{2N} {\norm{{\boldsymbol{\theta}^{(l)}}^*}}^2,\nonumber\\
&=&1-\frac{1}{2\lambda^2}J(L_m^{(l)},\tilde{\boldsymbol{\omega}}_m^{(l)}),
\end{eqnarray}
where the function $J(L_m^{(l)},\tilde{\boldsymbol{\omega}}_m^{(l)})$ is named as multiplex synchrony alignment function (MSAF) for the $l^{th}$ layer. Note that MSAF is a function of $L_m^{(l)}$ and $\tilde{\boldsymbol{\omega}}_m^{(l)}$ which depend on the Laplacians as well as dynamics of both the layers (see the equations (\ref{lm1}) - (\ref{omega_modified2})). Therefore, this MSAF is significantly different than its monolayer version in absence of frustration as reported in~\cite{Skardal_PRL2014}. Now, it is evident from the equation~(\ref{op_SAF}) that MSAF plays an important role for synchronization. As the value of MSAF increases, the system shows erotion of synchronization, while as $J(L_m^{(l)},\tilde{\boldsymbol{\omega}}_m^{(l)})\rightarrow 0$ the order parameter  $\mathrm{R}_{l}\rightarrow 1$, an enhancement in synchronization occurs.
The global order parameter of the system takes the form
\begin{eqnarray}\label{global_R}
\mathrm{R} &=& 1-\frac{1}{4N} {\norm{{\boldsymbol{\theta}^{(1)}}^*}}^2-\frac{1}{4N} {\norm{{\boldsymbol{\theta}^{(2)}}^*}}^2,\nonumber\\
&=&1-\frac{1}{4\lambda^2}J(L_m^{(1)},\tilde{\boldsymbol{\omega}}_m^{(1)})-\frac{1}{4\lambda^2}J(L_m^{(2)},\tilde{\boldsymbol{\omega}}_m^{(2)}).
\end{eqnarray}
It is now clear that the minimization of MSAF for each layer  will maximize the value of order parameters of individual layers as well as the global order parameter. 
Now,
\begin{eqnarray}\label{eqn5}
J(L_m^{(l)},\tilde{\omega}_m^{(l)})=\frac{1}{N}{\norm{{L_m^{(l)}}^{\dagger}\tilde{\omega}_m^{(l)}}}^2
\end{eqnarray}
${L_m^{(l)}}^{\dagger}$ can be expressed in terms of its eigen value and eigen vector as ${L_m^{(l)}}^{\dagger}= \sum e_j^{-1}v_j^{(l)}v_j^{(l)'}$ for all $e_j\neq 0$.
Hence $J(L_m^{(l)},\tilde{\boldsymbol{\omega}}_m^{(l)})$ takes the form
\begin{eqnarray}\label{eqn5_j}
J(L_m^{(l)},\tilde{\omega}_m^{(l)})=\frac{1}{N}\sum_j e_j^{-2} \langle v_j^{(l)}, \tilde{\omega}_m^{(l)}\rangle ^2.
\end{eqnarray}
If  ${\boldsymbol{\omega}}^{(l)}={\lbrace 0,0,\dots 0\rbrace}'$ then $\mathrm{R}^{(l)}$ will trivially be $1$. For non trivial solution we chose a standard deviation ($\sigma$) of the chosen frequency set such that $\sigma^2=\frac{\sum{\omega_i^{(l)}}^2}{N}$, where $\sigma$ is an arbitrary constant. Then we express $\omega^{(l)}$ as the linear combination of the eigenvectors of $L_m^{(l)}$ as $\boldsymbol{\omega}^{(l)}=\sum\alpha_j v_j^{(l)}$ where $\sum \alpha_j^2=\sigma^2N$.
Now choosing $\tilde{\boldsymbol{\omega}}_m^{(l)}=\sigma^2\sqrt{N}v_N^{(l)}$ we get $J(L_m^{(l)},\tilde{\boldsymbol{\omega}}_m^{(l)})\rightarrow 0$ and hence $R\rightarrow 1$.
Therefore,  the optimal frequency (after some calculation) can finally be obtained as
\begin{eqnarray}
{\boldsymbol{\omega}}^{(1)} &=& \sigma^2\sqrt{N}{L_m^{(2)}}^\dagger \lbrace v_N^{(1)} -v_N^{(2)} + \cos\beta L^{(1)}v_N^{(1)}\rbrace\nonumber\\
&+&\lambda d^{(1)} \sin\alpha, \label{opt_omegalayer1}\\
{\boldsymbol{\omega}}^{(2)}&=&\sigma^2\sqrt{N}{L_m^{(1)}}^\dagger \lbrace v_N^{(2)} -v_N^{(1)} + \cos\alpha L^{(2)}v_N^{(2)}\rbrace \nonumber\\
&+&\lambda d^{(2)} \sin\beta.\label{opt_omegalayer2}
\end{eqnarray}
As per the analytical scheme presented in this section, for the above choice of frequencies, the synchronization is expected to enhance substantially in the multiplex networks. In the next section we numerically verify the scheme and exploit it to achieve optimal as well as perfect synchronization in multiplex networks.

\section{Numerical verification}
For numerical verification of the scheme presented in the previous section to achieve optimal synchronization, we consider a heterogeneous multiplex network consisting of two layers each of which is a scale-free network of size $N=1000$ and exponent $\gamma=2.8$, while mean degrees of layer-I and layer-II are $\langle k_1 \rangle =8$ and $\langle k_2\rangle=12$ respectively. We then numerically simulate the system ~(\ref{eqn1})-~(\ref{eqn2}) using fourth order Runge-Kutta (RK4) scheme and compute the order parameters of each layer as well as the order parameter for the whole network as a function of coupling strength $\lambda$. 

\begin{figure*}
\includegraphics[height=!,width=0.95\textwidth]{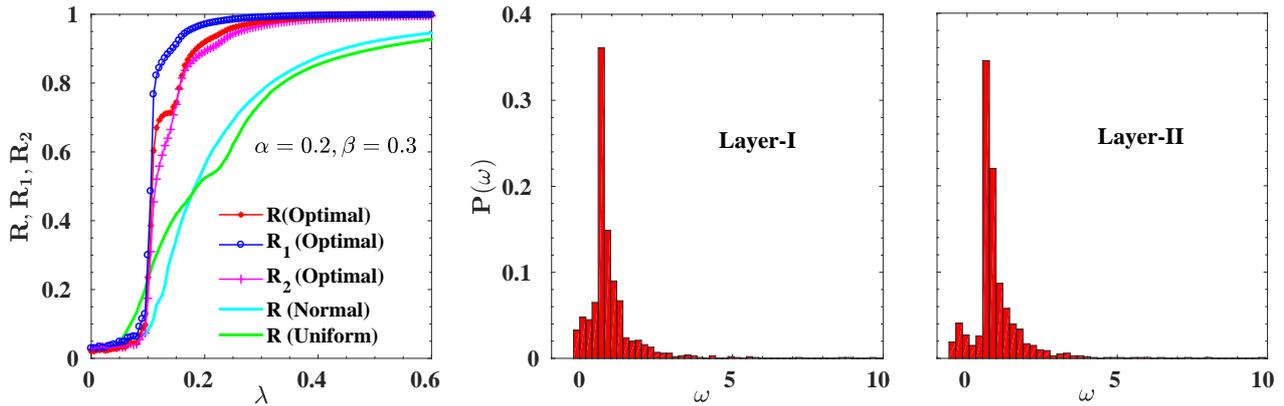}
\caption{{\bf Optimal Synchronization for nonzero $\alpha$ and $\beta$.} The multiplex network used here is same as the one used in figure~\ref{fig:1}. (a) Order parameters as a function of coupling strength $\lambda$ for optimal, uniform and normal frequency distributions.  (b) and (c) Optimal frequency distributions of the layer-I and layer-II respectively.} 
\label{fig:2}
\end{figure*}

{\it Optimal synchronization in non-frustrated multiplex networks.} First we use the scheme presented in the previous section to achieve optimal synchronization in non-frustrated multiplex networks ($\alpha=\beta=0$). In figure~\ref{fig:1}(a) we present the numerically computed order parameters of both the layers ($\mathrm{R_1}, \mathrm{R_2}$) as well as the global order parameter ($\mathrm{R}$) obtained from the simulation of the network using the frequency distribution obtained from the equations (\ref{opt_omegalayer1}- \ref{opt_omegalayer2}) as a function of the coupling strength $\lambda$ by taking $\sigma^2 = 1000$. Note that we have checked the numerical results for different values of $\sigma$ and always found that the chosen frequency help to achieve optimal synchronization in the multiplex network. However, larger values of $\sigma$ helps in increasing the range of the optimal frequency set. It also shows the variation of the global order parameter for Lorentzian distribution and uniform frequency distributions with $\lambda$. From the figure it is clear that the proposed frequency distribution helps the whole system to synchronize at lower coupling strength where as the other frequency distributions need much higher coupling strength to reach the synchronized state. 
The optimal frequency distributions of the layers I and II  are shown in Fig.\ \ref{fig:1}(b) and Fig.\ \ref{fig:1}(c) respectively. The bimodal nature of the distributions is apparent from the figures.  
 
\par {\it Optimal synchronization in frustrated multiplex networks.} 
Next we consider that both the layers of the multiplex network are phase phase frustrated ($\alpha, \beta  \neq 0$). Here we note that in presence of frustration,  the last terms of the equations (\ref{opt_omegalayer1}) and (\ref{opt_omegalayer2}) contain the coupling strength explicitly. Therefore, we calculate optimal frequencies around a desired  coupling strength 
$\lambda_{opt}=0.5$ for $\alpha = 0.2$ and $\beta = 0.3$ and numerically simulate the multiplex network. We also perform numerical simulation of the network using normal and uniform frequency distributions.  The order parameters computed from the simulation data in each case are shown in figure~\ref{fig:2}(a). From the figure it apparent that the system reaches to synchronized state at $\lambda = 0.5$ for the choice of optimal frequencies, while the system is far away from the synchronized state for other choice of frequencies. 

The optimal frequency distribution of the layers I and II are shown in Fig.\ \ref{fig:2} (b) and (c) respectively. In presence of frustration,  the optimal frequencies show unimodal distribution as opposed to bimodal distribution obtained for non-frustrated dynamics (Fig.\ \ref{fig:1} (b) and (c)). It appears from the expressions (\ref{opt_omegalayer1}) and (\ref{opt_omegalayer2}) of optimal frequency that the last terms of the expressions play a crucial role in determining the nature of the distributions. This term is absent in the non-frustrated case. We would like to mention here we have performed similar exercise with scale-free (SF) network in layer-I and Erd\"{o}s-R\'{e}nyi (ER) in layer-II, and also with ER networks in both the layers. In all the cases the presented scheme provided the optimal synchronization in the multiplex. Note that, if we select the optimal frequency for higher coupling strength then ($\lambda_{opt} > 0.5$) the coupling strength will take higher value to synchronize globally.  However, these frequencies can not produce perfect synchronization ($\mathrm{R}=1$). 
 
\par In the next section we consider multiplex network with both frustrated and non frustrated dynamics in the layers. Then using our analytical scheme we investigate the possibility  of achieving perfect synchronization ($\mathrm{R} = 1$) in the multiplex network. 

\begin{figure}[h]
\includegraphics[height=!,width=8.8cm]{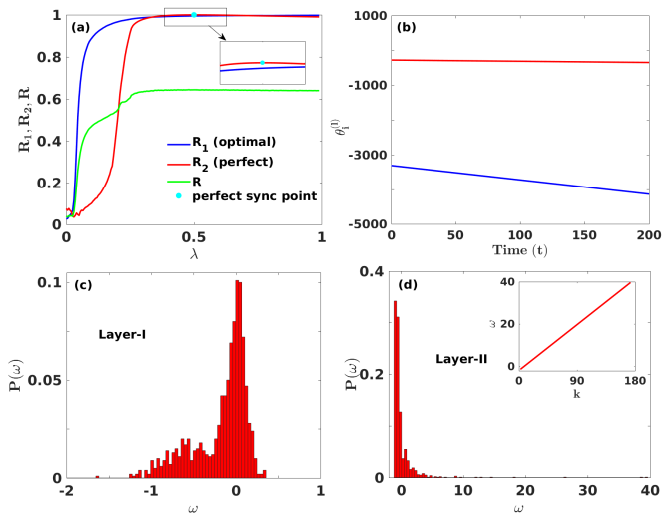}
\caption{{\bf Optimal and Perfect synchronization in different layers.} Same multiplex network used here as the previous figures with $\alpha = 0$ and $\beta = 0.5$ (a) Order parameters as a function of coupling strength $\lambda$. (b) Time evolution of the phase values of layer-I (blue curve) and layer - II (red curves). (c) Optimal frequency distribution of layer - I. (d) Frequency ($\mathbf(\omega)$ as a function of degree ($\mathrm{k}$) of the nodes in layer - II. }
\label{fig:3}
\end{figure}
\begin{figure*}
\includegraphics[height=!,width=0.95\textwidth]{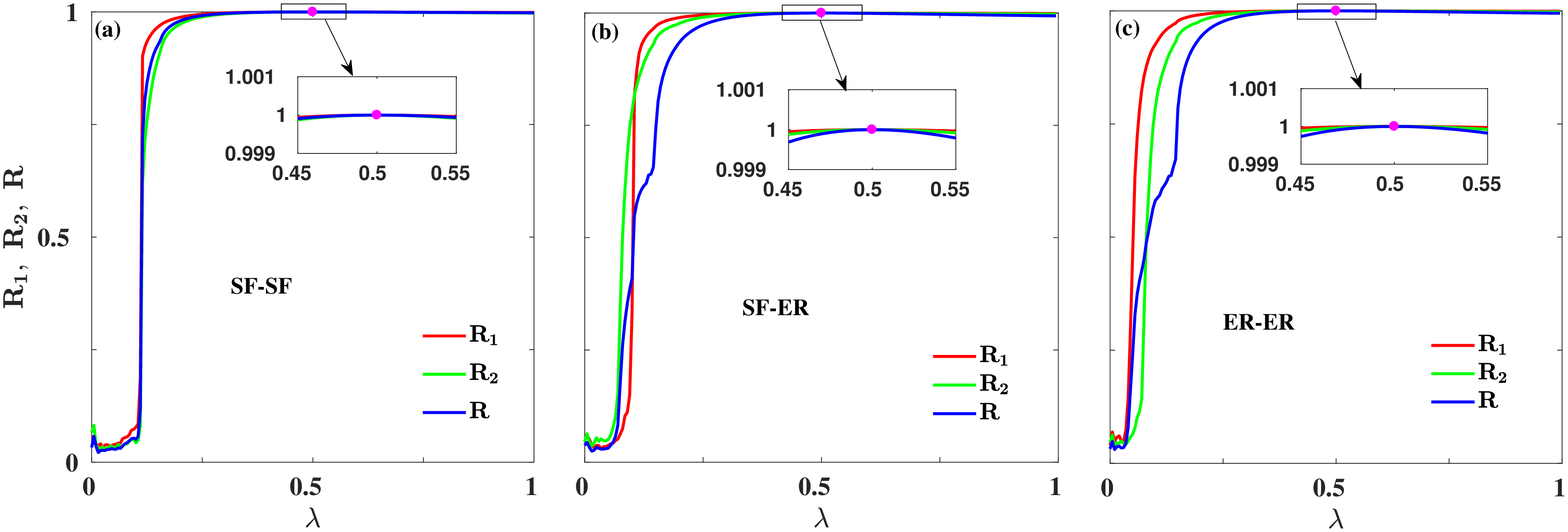}
\caption{{\bf Perfect synchronization in different layers.} Variation of the order parameters for individual layers ($\mathrm{R_1}$) and ($\mathrm{R_2}$) as well as the global order parameter $\mathrm{R}$ ( red, green and blue curves respectively) as a function of the coupling strength ($\lambda$) for  $\alpha = 0.2$ and $\beta = 0.3$. Different networks of size $N=1000$ are considered in different layers. (a) Both layers are scale-free (SF) with exponent $\gamma = 2.8$ (For layer-I and II average degrees are $\langle k\rangle=12$ and $\langle k\rangle=8$ respectively). (b) Layer-I is scale-free with $\gamma = 2.8$ and $\langle k\rangle=12$ and layer-II is Erd\"{o}s-R\'{e}nyi with $\langle k\rangle=10$. (c) Both the layers are ER networks (For layer-I, $\langle k\rangle=12$ and for layer-II, $\langle k\rangle=10$). In all three cases the perfect synchronization is achieved at $\lambda_p=0.5$ (magenta dot). Zoomed views near the targeted points clearly show the achievement of perfect synchronization.}
\label{fig:4}
\end{figure*}
\section{Perfect and optimal synchronization}
It has been reported earlier\cite{Kundu_EPL2018} that by appropriate choice of nodal frequencies, perfect synchronization ($\mathrm{R} = 1$ can be achieved in a frustrated network (monolayer) of phase oscillators for given coupling strength and phase frustration.  Here, in case of multiplex networks, if the nonzero phase frustrations exists in the system, we find that it may exhibit  perfect synchronization state where individual layer will attain perfect synchronization at a particular coupling strength with given values of $\alpha$ and $\beta$.
To achieve perfect synchronization state in each layer, the order parameter in the layers must reach $\mathrm{R}_l=1$ and consequently from equation\ (\ref{op_SAF}) we have $J(L_m^{(l)},\tilde{\boldsymbol{\omega}}_m^{(l)})=0$ ($l = 1,2$). The simplest choice which satisfy this condition is $\tilde{\boldsymbol{\omega}}_m^{(l)}=0$ ($l = 1,2$). To maintain this condition we may set  $\tilde{\boldsymbol{\omega}}^{(1)}$ and $\tilde{\boldsymbol{\omega}}^{(2)}$ to zero simultaneously (see equations (\ref{omega_modified1}-\ref{omega_modified2})). Therefore,  the frequency set of each layer ${\boldsymbol{\omega_p}}^{(l)}$ is given by
\begin{eqnarray}
{\boldsymbol{\omega_p}}^{(1)}=\lambda_p\sin\alpha \boldsymbol{d}^{(1)}, \label{perfect_omega1}\\
{\boldsymbol{\omega_p}}^{(2)}=\lambda_p\sin\beta \boldsymbol{d}^{(2)}, \label{perfect_omega2}
\end{eqnarray}
where $\lambda_p$ is the coupling strength at which we predict perfect synchronization in the multiplex network for given $\alpha$ and $\beta$. From the equations~(\ref{perfect_omega1}) and~(\ref{perfect_omega2}) we find that the set of frequencies to achieve perfect synchronization in a layer of the multiplex network entirely depends on its own degree distribution (linearly related to its own degree) and frustration. In the following we consider multiplexes with frustrated and non frustrated dynamics in the layers and use the analytical scheme derived above to achieve perfect and optimal synchronizations in different layers.

{\it Multiplex with mixed dynamics.} For numerical verification, first we consider mixed dynamics in the multiplex. Layer - I is governed by non frustrated dynamics, while the layer - II is governed by frustrated dynamics and we set $\alpha = 0$ and $\beta = 0.5$. We assign frequencies in layer - I (non frustrated) and layer - II (frustrated) as obtained from the equations~(\ref{opt_omegalayer1}) and -\ref{perfect_omega2}) respectively in order to achieve optimal synchronization in layer-I and perfect synchronization in layer - II at a targeted coupling strength $\lambda_p = 0.5$.  We then simulate the entire network for different coupling strengths using RK4. Figure~\ref{fig:3}(a) shows the variation of the order parameters of the individual layers ($\mathrm{R_1}, \mathrm{R_2}$, blue and red curve) as well as the global order parameter ($\mathrm{R}$, green curve) as a function of coupling strength as obtained from the numerical simulations. The figure shows that the frequencies derived from the equations~(\ref{opt_omegalayer1}) and -\ref{perfect_omega2}) fulfill the expectation of achieving optimal and  perfect synchronization in individual layers. The perfect synchronization can not be established in non-frustrated layer-I  but order parameter ($\mathrm{R_2}$) of  layer-II touches 
 $1$ at $\lambda_p=0.5$ (cyan dot, blue and red curves shown in the inset of  Fig. \ref{fig:3} (a)). 

Interesting to note that the global order parameter ($\mathrm{R}$) does not reach the high level of  synchronization. The reason is that the layers are separately synchronized to different phases and  their synchronized  phase values maintain a growing distance between them (Fig.\ \ref{fig:3} (b)). As expected the frequencies of second layer is linearly related with degree (Fig.\ \ref{fig:3} (d)), and frequencies of first layer do not follow any specific relation with the degree of the considered network (Fig.\ \ref{fig:3} (c)).

{\it Multiplex with frustrated dynamics.} Finally we consider frustrated dynamics in both the layers of the multiplex network. We choose $\alpha = 0.2$ and $\beta = 0.3$ and derive the frequency distributions from the equations~(\ref{perfect_omega1}) and (\ref{perfect_omega2}) which we assign to the layers I and II respectively to achieve perfect synchronization in both the layers at the targeted coupling strength $\lambda_p = 0.5$. Figure~\ref{fig:4} shows the order parameters as function of the coupling strength as obtained from the numerical simulations. The figure clearly demonstrates the achievement of perfect synchronization in the network at a preassigned coupling strength $\lambda_p$ for given frustration parameters. Therefore, the analytically derived frequency distributions for both the layers are found to work well for the achievement of optimal as well as perfect synchronization in the layers at a desired coupling strength.  

\section{Conclusion}
A general mathematical framework is developed in this paper to derive the natural frequencies of the nodes of  multiplex networks which can ensure high level of synchronization at considerably lower coupling strength.  The framework is based on the derivation of a multiplex syncrony alignment function (MSAF). The analysis of the MSAF using the theory of linear algebra provide a way to analytically determine a set of natural frequencies for the network which ensure optimal synchronization in the layers of the network. It is found that the analytically derived natural frequencies involve both structural and dynamical information of the phase frustrated multiplex network. We have shown that optimal frequency of a layer depends on the leading eigen vector of the underlying  network Laplacian and the pseudo inverse operator of both the layers. 
Further we have identified the condition for perfect and optimal synchronization in  multiplex networks in which one layer is frustrated and the other layer is non frustrated. 

\section{Acknowledgements}PK acknowledges support from  DST, India under the DST-INSPIRE scheme (Code: IF140880). CH is supported by INSPIRE-Faculty grant (Code: IFA17-PH193).


\begin{thebibliography}{10}
\expandafter\ifx\csname url\endcsname\relax\def\url#1{\texttt{#1}}\fi

\bibitem{Pikovsky_synchronization_book}
\Name{Pikovsky A., Rosenblum M., Kurths J. \and Kurths J.}
  \Book{Synchronization: a universal concept in nonlinear sciences} Vol.~12
  (Cambridge university press) 2003.

\bibitem{Arenas_PhysReport2008}
\Name{Arenas A., D{\'\i}az-Guilera A., Kurths J., Moreno Y. \and Zhou C.}
  \REVIEW{Physics reports}{469}{2008}{93}.

\bibitem{Strogatz_synchronization_book}
\Name{Strogatz S.} \Book{Sync: The emerging science of spontaneous order}
  (Penguin UK) 2004.

\bibitem{Motter_NatPhys2013}
\Name{Motter A.~E., Myers S.~A., Anghel M. \and Nishikawa T.} \REVIEW{Nature
  Physics}{9}{2013}{191}.

\bibitem{Belykh_PRL2005}
\Name{Belykh I., de~Lange E. \and Hasler M.} \REVIEW{Physical review
  letters}{94}{2005}{188101}.

\bibitem{Dorogovtsev_RMP2008}
\Name{Dorogovtsev S.~N., Goltsev A.~V. \and Mendes J.~F.} \REVIEW{Reviews of
  Modern Physics}{80}{2008}{1275}.

\bibitem{Albert_RMP2002}
\Name{Albert R. \and Barab{\'a}si A.-L.} \REVIEW{Reviews of modern
  physics}{74}{2002}{47}.

\bibitem{Cohen_complex_book}
\Name{Cohen R. \and Havlin S.} \Book{Complex networks: structure, robustness
  and function} (Cambridge university press) 2010.

\bibitem{Barzel_NatPhys2013}
\Name{Barzel B. \and Barab{\'a}si A.-L.} \REVIEW{Nature physics}{9}{2013}{673}.

\bibitem{Hens_NatPhysics2019}
\Name{Hens C., Harush U., Haber S., Cohen R. \and Barzel B.} \REVIEW{Nature
  Physics}{15}{2019}{403}.

\bibitem{Ichinomiya_PRE2004}
\Name{Ichinomiya T.} \REVIEW{Physical Review E}{70}{2004}{026116}.

\bibitem{Restrepo_PRE2005}
\Name{Restrepo J.~G., Ott E. \and Hunt B.~R.} \REVIEW{Physical Review
  E}{71}{2005}{036151}.

\bibitem{Arenas_PRL2006}
\Name{Arenas A., D{\'\i}az-Guilera A. \and P{\'e}rez-Vicente C.~J.}
  \REVIEW{Physical review letters}{96}{2006}{114102}.

\bibitem{Jesus_PRL2007}
\Name{G{\'o}mez-Gardenes J., Moreno Y. \and Arenas A.} \REVIEW{Physical review
  letters}{98}{2007}{034101}.

\bibitem{Skardal_PRL2014}
\Name{Skardal P.~S., Taylor D. \and Sun J.} \REVIEW{Physical review
  letters}{113}{2014}{144101}.

\bibitem{Boccaletti_PhysReport2016}
\Name{Boccaletti S., Almendral J., Guan S., Leyva I., Liu Z., Sendi{\~n}a-Nadal
  I., Wang Z. \and Zou Y.} \REVIEW{Physics Reports}{660}{2016}{1}.

\bibitem{Domenico_NatPhys2016}
\Name{De~Domenico M., Granell C., Porter M.~A. \and Arenas A.} \REVIEW{Nature
  Physics}{12}{2016}{901}.

\bibitem{Danziger_NatPhys2019}
\Name{Danziger M.~M., Bonamassa I., Boccaletti S. \and Havlin S.}
  \REVIEW{Nature Physics}{15}{2019}{178}.

\bibitem{Jalan_PRE2019}
\Name{Jalan S., Rathore V., Kachhvah A.~D. \and Yadav A.} \REVIEW{Physical
  Review E}{99}{2019}{062305}.

\bibitem{Kachhvah_EPL2017}
\Name{Kachhvah A.~D. \and Jalan S.} \REVIEW{EPL (Europhysics
  Letters)}{119}{2017}{60005}.

\bibitem{Khanra_PRE2018}
\Name{Khanra P., Kundu P., Hens C. \and Pal P.} \REVIEW{Physical Review
  E}{98}{2018}{052315}.

\bibitem{Nicosia_PRL2017}
\Name{Nicosia V., Skardal P.~S., Arenas A. \and Latora V.} \REVIEW{Physical
  review letters}{118}{2017}{138302}.

\bibitem{Panos_PRE2019}
\Name{Soriano-Pa{\~n}os D., Guo Q., Latora V. \and G{\'o}mez-Garde{\~n}es J.}
  \REVIEW{Physical Review E}{99}{2019}{062311}.

\bibitem{Jesus_PRL2011}
\Name{G{\'o}mez-Gardenes J., G{\'o}mez S., Arenas A. \and Moreno Y.}
  \REVIEW{Physical review letters}{106}{2011}{128701}.

\bibitem{Kundu_EPL2018}
\Name{Kundu P., Hens C., Barzel B. \and Pal P.} \REVIEW{EPL (Europhysics
  Letters)}{120}{2018}{40002}.

\bibitem{Brede_PRE2016}
\Name{Brede M. \and Kalloniatis A.~C.} \REVIEW{Physical Review
  E}{93}{2016}{062315}.

\bibitem{Kuramoto_chemical_book}
\Name{Kuramoto Y.} \Book{Chemical oscillations, waves, and turbulence} (Courier
  Corporation) 2003.

\bibitem{Lohe_Automatica2015}
\Name{Lohe M.} \REVIEW{Automatica}{54}{2015}{114}.

\bibitem{Wiesenfeld_PRE1998}
\Name{Wiesenfeld K., Colet P. \and Strogatz S.~H.} \REVIEW{Physical Review
  E}{57}{1998}{1563}.

\bibitem{Dorfler_SIAM2012}
\Name{Dorfler F. \and Bullo F.} \REVIEW{SIAM Journal on Control and
  Optimization}{50}{2012}{1616}.

\bibitem{Mertens_PRE2011}
\Name{Mertens D. \and Weaver R.} \REVIEW{Physical Review E}{83}{2011}{046221}.

\bibitem{Omelchenko_PRL2012}
\Name{Omel’chenko E. \and Wolfrum M.} \REVIEW{Physical review
  letters}{109}{2012}{164101}.

\bibitem{Sakaguchi_PTP1986}
\Name{Sakaguchi H. \and Kuramoto Y.} \REVIEW{Progress of Theoretical
  Physics}{76}{1986}{576}.

\bibitem{Kundu_PRE2017}
\Name{Kundu P., Khanra P., Hens C. \and Pal P.} \REVIEW{Physical Review
  E}{96}{2017}{052216}.

\bibitem{Kundu_Chaos2019}
\Name{Kundu P. \and Pal P.} \REVIEW{Chaos: An Interdisciplinary Journal of
  Nonlinear Science}{29}{2019}{013123}.

\end{thebibliography}


\end{document}